\documentclass[aps,pre,twocolumn,floatfix,nofootinbib]{revtex4}
\usepackage{amssymb,amsmath,amsthm,mathrsfs,latexsym}
\usepackage[sort&compress]{natbib}
\usepackage{graphicx}
\usepackage{epsfig}
\usepackage{epstopdf}
\usepackage{color}

\begin{document}

\title{Inter-state switching in stochastic gene expression: \\ Exact solution, an adiabatic limit and oscillations in molecular distributions}

\author{K.G. Petrosyan}
\affiliation{Department of Physics, National Dong Hwa University, Hualien 974, Taiwan}
\affiliation{Institute of Physics, Academia Sinica, Nankang, Taipei 11529, Taiwan}
\date{\today}

\begin{abstract}
We consider the stochastic gene expression process with inter-state flip-flops. An exact steady-state solution to the master equation is calculated.
One of the main goals in this paper is to investigate whether the probability distribution of gene copies contains even-odd number oscillations. 
A master equation previously derived in the adiabatic limit of fast switching by Kepler and Elston \cite{kepler}  suggests that the oscillations 
should be present. However our analysis demonstrates that the oscillations do not happen not only in the adiabatic case  but they are entirely absent.
We discuss the adiabatic approximation in detail. The other goal is to establish the master equation that takes into account external fluctuations
that is similar to the master equation in the adiabatic approximation. The equation allows even-odd oscillations. The reason the behaviour occurs 
is an underlying interference of Poisson and Gaussian processes.  The master equation contains an extra term that describes the gene copy number 
unconventional diffusion and is responsible for the oscillations. We also point out to a similar phenomenon in quantum physics.
\end{abstract}

\maketitle

{\it Introduction. - }Stochastic gene expression in the process of transcriptional regulation was studied in \cite{kepler} using
the (bio)chemical master equation which takes into account switching between two steady states of gene copy production. The model
presented in \cite{kepler} had been then studied by various researchers (see, e.g, \cite{sugar,duncan}) and describes a number of phenomena
with noise-induced multistability being a rather remarkable one. The stochastic gene expression and regulation remains an active area
of research \cite{eldar,genewei,paulsson,golding,shalek,oudenaarden,bressloff} with modern experimental advancements making it possible 
to measure mRNA and protein copy abundances with single-molecule sensitivity \cite{shalek,skinner,skinner2,yanagida,larsson}.

The master equation exploited by Kepler and Elston \cite{kepler} for the stochastic process of transcriptional regulation without feedback
reads as
\begin{widetext}
\begin{eqnarray}
\frac{dP_n^0}{dt}=\epsilon_0 (P_{n-1}^0-P_n^0)+\frac{1}{\tau}[(n+1)P_{n+1}^0-nP_n^0]+K(k_1P_n^1-k_0P_n^0) \nonumber \\
\frac{dP_n^1}{dt}=\epsilon_1 (P_{n-1}^1-P_n^1)+\frac{1}{\tau}[(n+1)P_{n+1}^1-nP_n^1]+K(k_0P_n^0-k_1P_n^1)
\label{2sme}
\end{eqnarray}
\end{widetext}
where $P_n^0$ and $P_n^1$ are the probability distributions for having $n$ particles produced while in the state $0$ or $1$, respectively;
$\epsilon_{0,1}$ are the production rates; $\frac{1}{\tau}$ is the degradation rate taken equal for both states. $k_0$ and $k_1$ are
inter-state switching rates and it is required that $k_0+k_1=1$. $K$ sets the time scale for the flip-flop process.
Among several interesting findings in \cite{kepler} there is the following equation for the marginal probability distribution $P_n=P_n^0+P_n^1$
in the adiabatic limit of $k_1\epsilon_0+k_0\epsilon_1 \ll K$ \cite{kepler}
\begin{widetext}
\begin{eqnarray}
\frac{dP_n}{dt}=(k_1\epsilon_0+k_0\epsilon_1)(P_{n-1}-P_n)+\frac{1}{\tau}[(n+1)P_{n+1}-nP_n]+\frac{k_0k_1}{K}(\epsilon_0-\epsilon_1)^2(P_{n-2}-2P_{n-1}+P_n)
\label{adiame}
\end{eqnarray}
\end{widetext}
We have noticed that this equation has the following steady-state solution
\begin{widetext}
\begin{eqnarray}
P_n=\frac{(i\sigma\sqrt{\frac{\tau}{2}})^n}{n!}H_n\left(i\sigma\sqrt{\frac{\tau}{2}}\left(1-\frac{\bar{\epsilon}}{\sigma^2\tau}\right)\right)\cdot\exp\left[-\bar{\epsilon}+\frac{1}{2}\sigma^2\tau\right]
\label{solution}
\end{eqnarray}
\end{widetext}
where $\bar{\epsilon}=k_1\epsilon_0+k_0\epsilon_1$ and $\sigma^2=\frac{k_0k_1}{K}(\epsilon_0-\epsilon_1)^2$. 

This can be checked via substituting the expression into the master equation and making use of the recurrence relation $H_{n+1}(x)=2xH_n(x)-2nH_{n-1}(x)$ for the Hermite polynomials \cite{abramowitz}. This is exactly the same probability distribution as obtained in \cite{oscillations} for the case of extrinsic noise when replacing
the production rate $\epsilon$ with $\bar{\epsilon}+\sigma\xi(t)$ with $\xi(t)$ being the delta-correlated noise source with zero mean $<\xi(t)>=0$ , $<\xi(t)\xi(t')>=\delta(t-t')$  and $\sigma^2$ the intensity of the noise. 
In the Eq.(\ref{adiame}) $k_1\epsilon_0+k_0\epsilon_1$ corresponds to $\epsilon$ and $\frac{k_0k_1}{K}(\epsilon_0-\epsilon_1)^2$ corresponds to $\sigma^2$ as mentioned above.

\begin{figure}
\includegraphics[width=1.05\linewidth,angle=0]{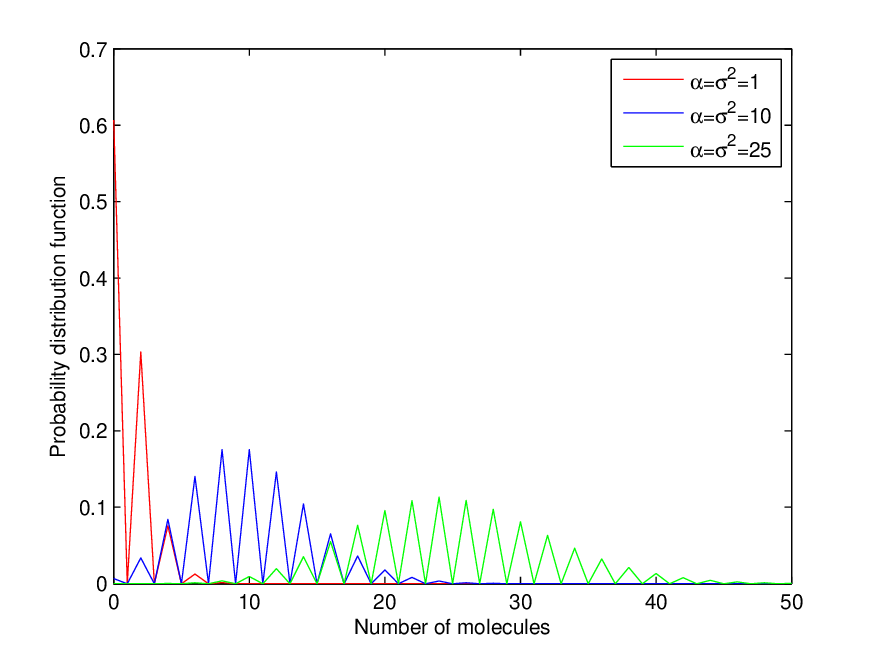}
\caption{(Color online) Probability distributions (\ref{solution}) for $\bar\epsilon=\sigma^2$, $\tau=1$.}
\label{fig1}
\end{figure}

\begin{figure}
\includegraphics[width=1.05\linewidth,angle=0]{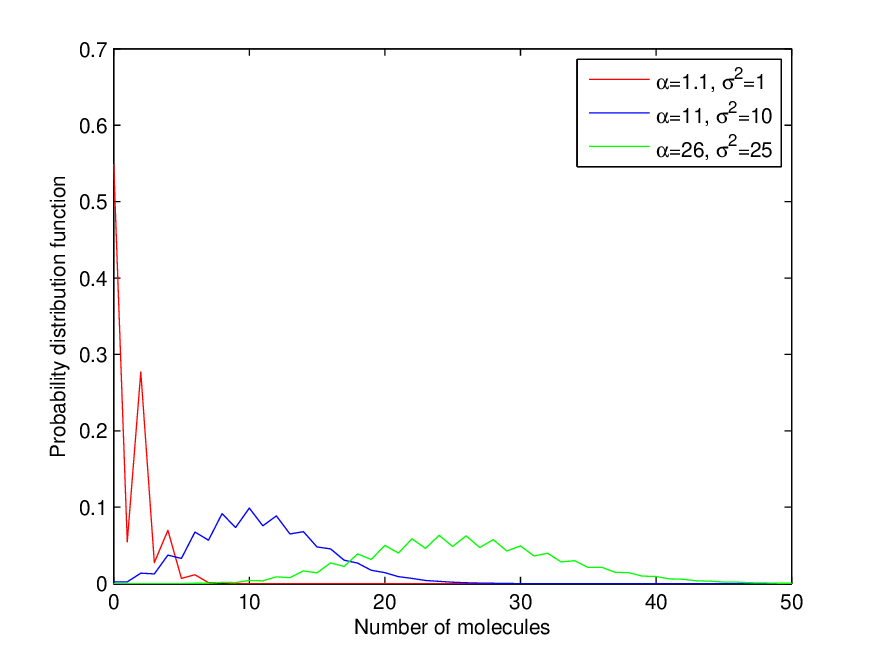}
\caption{(Color online) Probability distributions (\ref{solution}) for $\bar\epsilon\simeq\sigma^2$, $\tau=1$.}
\label{fig2}
\end{figure}

Meanwhile the probability distribution function (\ref{solution}) has a remarkable feature -- the even/odd oscillations studied in \cite{oscillations,pseudogenes,multistability,doubly}. 
The presence of even-odd oscillations in the number of particles for the distribution (\ref{solution}) is illustrated in Figs.(\ref{fig1}) and (\ref{fig2}). Thus a question arises whether the oscillations are present in
the model of Kepler and Elston \cite{kepler} for the stochastic process of transcriptional regulation without feedback. The answer is they are absent in that particular model. Not just in the adiabatic limit but they are entirely absent. 
We will proceed with calculations and will explain what exactly was done in the paper \cite{kepler} and point to an error as well as demonstrate what kind of approximation would lead to the same results as in \cite{kepler}.

The other goal is to establish a master equation in the form similar to that of (\ref{adiame})
\begin{eqnarray}
\frac{dP_n}{dt}=\hat L P_n+\sigma^2(P_{n-2}-2P_{n-1}+P_n)
\end{eqnarray}
with $\hat L P_n$ being the conventional (standard) part of the (bio)chemical master equation. The master equation contains an extra term  that takes into account external fluctuations \cite{oscillations}
and describes the unconventional diffusion of the number of particles that is responsible for the oscillations. 
The reason the behaviour occurs is an underlying convolution of Poisson and Gaussian processes as explained below. 

We have also noticed that there are oscillations in molecular abundances as well as bimodality (bistability) measured in a number of experiments \cite{skinner,skinner2,yanagida,larsson} and will briefly discuss
if our findings are related to the data.

In order to investigate the system we will apply the Poisson representation approach \cite{statphys,gardiner} that allows to derive Fokker-Planck equation for a quasiprobability function from the master equation
without any approximation (no need for, e.g., systeim-size expansion). The method gives exact results for arbitrary number of particles (small or large) as well as for time-varying parameters of the set of
molecular reactions such as production rates \cite{statphys,gardiner,drummond,thomas,sugar,seafr,gallegati,burnett,schnoerr}. 

\begin{figure}
	\includegraphics[width=1.05\linewidth,angle=0]{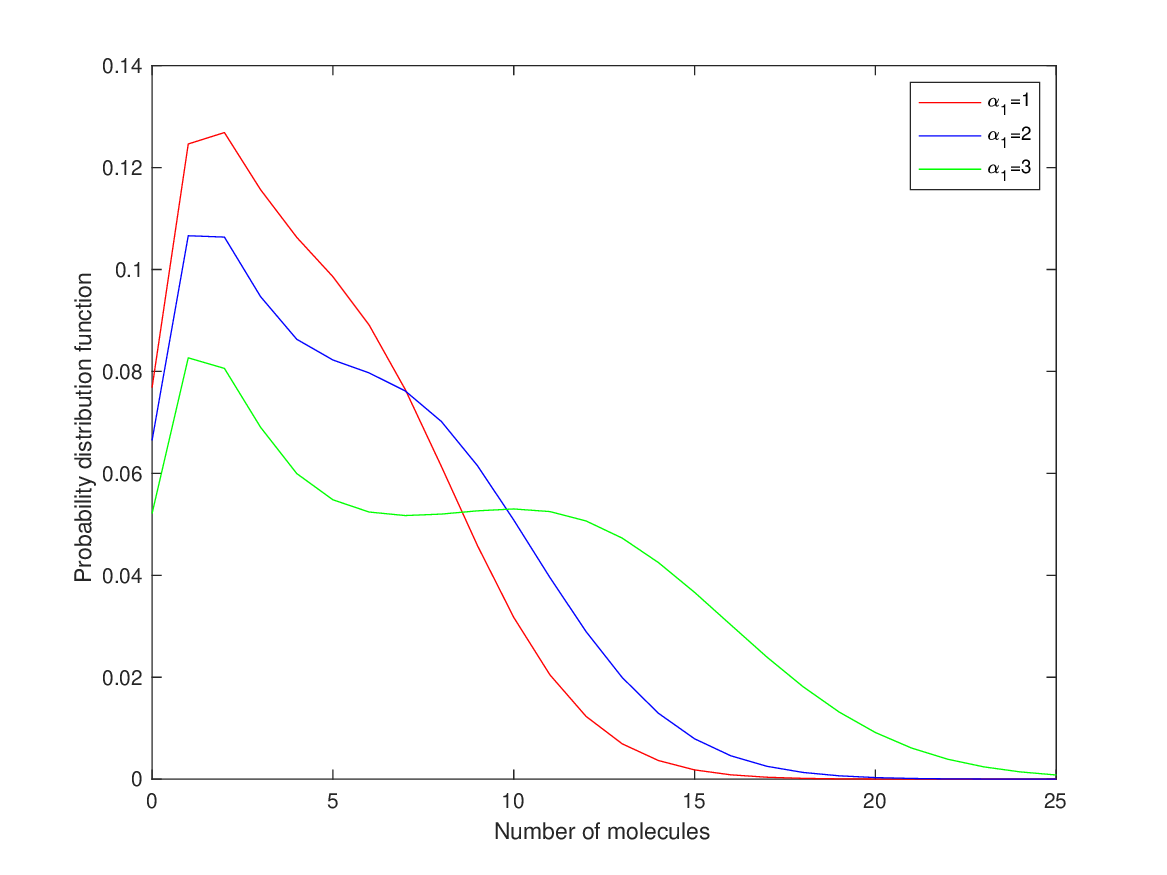}
	\caption{(Color online) Probability distributions for the following set of parameters: $K=1$; $k_0=k_1=0.5$, $\alpha_0=0.1$.}
	\label{fig3}
\end{figure}

\begin{figure}
	\includegraphics[width=1.05\linewidth,angle=0]{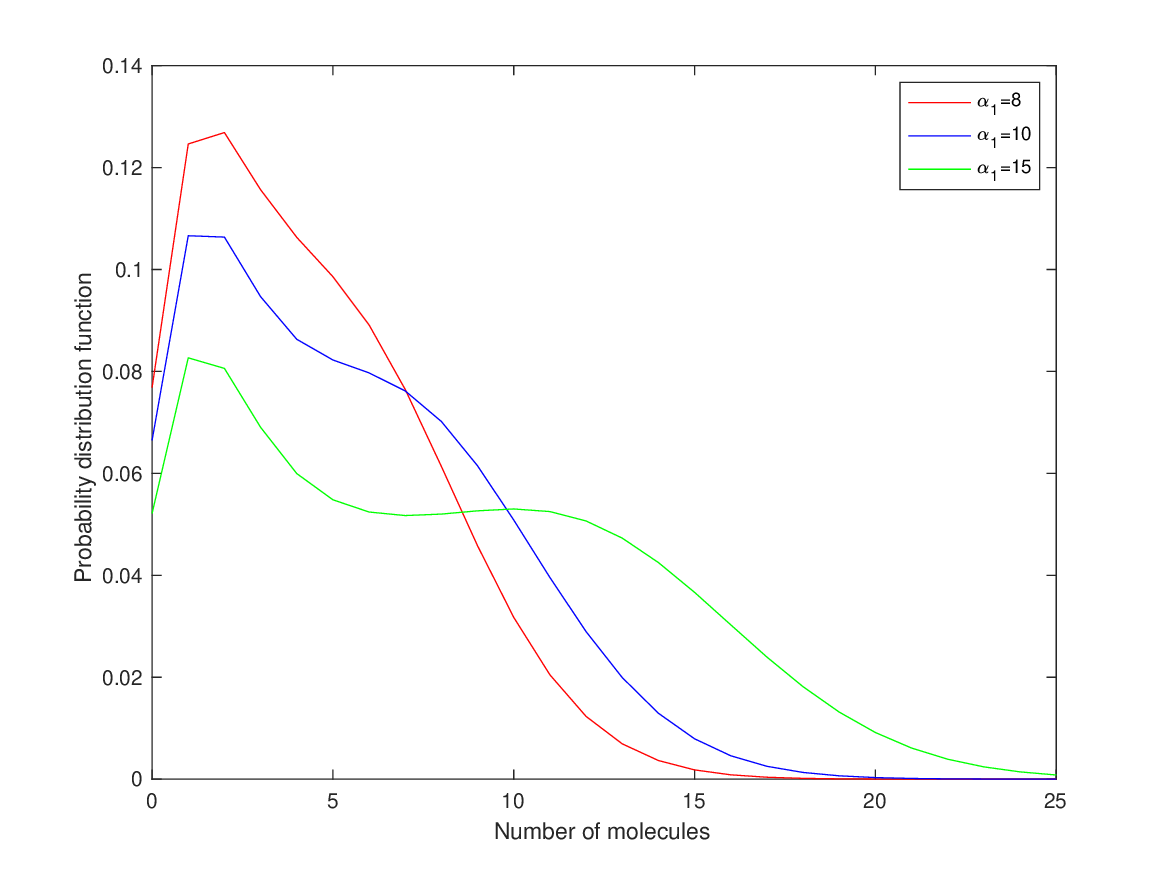}
	\caption{(Color online) Probability distributions for the following set of parameters: $K=1$, $k_0=k_1=0.5$, $\alpha_0=1$.}
	\label{fig4}
\end{figure}

{\it Exact solution and the large $K$ limit. - }Let us now find a stationary solution to the 2-state master equation (\ref{2sme}). Using the Poisson representation approach
 \cite{statphys,gardiner,drummond,thomas,sugar,seafr,gallegati,burnett,schnoerr}  that is substituing
\begin{eqnarray}
P_n^{0,1}=\int d\alpha \frac{\alpha^n}{n!}e^{-\alpha} f_{0,1}(\alpha) \nonumber
\end{eqnarray}
into the master equation (\ref{2sme}) and making corresponding transformations \cite{gardiner} we obtain the following equations for the quasiprobability functions $f_{0,1}(\alpha)$
\begin{eqnarray}
\frac{\partial}{\partial t} f_0(\alpha) = -\frac{\partial}{\partial \alpha} \left[\frac{1}{\tau}(\alpha_0-\alpha)f_0(\alpha)\right] + K[k_1f_1(\alpha)-k_0f_0(\alpha)] \nonumber \\
\frac{\partial}{\partial t} f_1(\alpha) = -\frac{\partial}{\partial \alpha} \left[\frac{1}{\tau}(\alpha_1-\alpha)f_1(\alpha)\right] + K[k_0f_0(\alpha)-k_1f_1(\alpha)] \nonumber
\end{eqnarray}
It is easy to check that the following expressions are the steady-state solutions (we define $\alpha_{0,1}=\epsilon_{0,1}\tau$ and will assume that $\alpha_0<\alpha_1$). 
\begin{eqnarray}
f_0(\alpha)=\frac{C}{\alpha-\alpha_0}(\alpha-\alpha_0)^{Kk_0\tau}(\alpha_1-\alpha)^{Kk_1\tau} \nonumber \\
f_1(\alpha)=\frac{C}{\alpha_1-\alpha}(\alpha_1-\alpha)^{Kk_1\tau}(\alpha-\epsilon_0)^{Kk_0\tau} \nonumber
\end{eqnarray}
with the same normalization constant $C$. The solutions written in this form show that in case there is no switching $K=0$ one can use the complex Poisson representation technique \cite{gardiner} and obtain Poisson distributions for both $P_n^0$ and $P_n^1$ with the mean numbers $\alpha_0$ and $\alpha_1$.

The quasiprobability distribution function $f(\alpha)=f_0(\alpha)+f_1(\alpha)$ which corresponds to $P_n=P_n^0+P_n^1$ can be expressed as
\begin{eqnarray}
f(\alpha)=C(\epsilon_1-\alpha_0)(\alpha-\epsilon_0)^{Kk_0\tau-1}(\epsilon_1-\alpha)^{Kk_1\tau-1}
\label{qpdf}
\end{eqnarray}
The probability distribution for the number of particles becomes
\begin{eqnarray}
P_n=C_1\int_{\alpha_0}^{\alpha_1} d\alpha \cdot \frac{\alpha^n}{n!}e^{-\alpha}\cdot(\alpha-\alpha_0)^{Kk_0\tau-1}(\alpha_1-\alpha)^{Kk_1\tau-1}
\label{pdf}
\end{eqnarray}
where $C_1=C(\alpha_1-\alpha_0)$. 
Summing up both sides of (\ref{pdf}) for all values \(n\geqslant 0\) and recalling that sum of all probabilities on the left side will be equal to \(1\) while exchanging the order of summation and integration on the right we obtain the identy
\[
1=C_1\int_{\alpha_0}^{\alpha_1}  (\alpha-\alpha_0)^{Kk_0\tau-1}(\alpha_1-\alpha)^{Kk_1\tau-1}\,d\alpha
\]
or making a change of variables \(\alpha=(\alpha_1-\alpha_0)u+\alpha_0\)
\[
1=C_1(\alpha_1-\alpha_0)^{K\tau-1}\int_{0}^{1}  u^{Kk_0\tau-1}(1-u)^{Kk_1\tau-1}\,du
\]
Hence recognizing the integral on the right as Beta function \cite{abramowitz} which is expressible in terms of Gamma functions \cite{abramowitz} we obtain the following expression for the constant
\[
C_1=(\alpha_1-\alpha_0)^{1-K\tau}\frac{\Gamma(Kk_0\tau)\Gamma(Kk_1\tau)}{\Gamma(K\tau)}
\]
In a similar way we can express the integral in (\ref{pdf}) through Kummer's \(M\) function defined as  \cite{abramowitz} 
\[
M(a,b,z)=\frac{\Gamma(b)}{\Gamma(a)\Gamma(b-a)}\int_{0}^{1}e^{zu}u^{a-1}(1-u)^{b-a-1}\,du
\]
Indeed, making the same change of variables \(\alpha=(\alpha_1-\alpha_0)u+\alpha_0\) we obtain
\[
\begin{split}
P_n&=\frac{C_1}{n!}(\alpha_1-\alpha_0)^{K\tau-1}e^{\alpha_0}
\\
\times&\int_{0}^{1} ((\alpha_1-\alpha_0)u+\alpha_0)^n e^{(\alpha_1-\alpha_0)u} u^{Kk_0\tau-1}(1-u)^{Kk_1\tau-1}\,du
\end{split}
\]
Applying now binomial theorem, recalling the definition of the Kummers function and substituting the expression for $C_1$ we get 
\[
\begin{split}
P_n&=\frac{e^{\alpha_0}}{n!}\frac{\Gamma(Kk_0\tau)\Gamma(Kk_1\tau)}{\Gamma(K\tau)}
\\
\times&\sum_{k=0}^{n}\Biggl(\binom{n}{k}(\alpha_1-\alpha_0)^{k}\alpha_0^{n-k}
M(Kk_0\tau+k,Kk_1\tau,\alpha_1-\alpha_0)
\\
&\times\frac{\Gamma(Kk_0\tau+k)\Gamma(Kk_1\tau)}{\Gamma(K\tau+k)}\Biggr)
\end{split}
\]

This exact solution is not novel. It is a generalized ($\alpha_0 > 0$) linear case of an exact solution obtained in \cite{sugar}. Our goal is to find out if there are any even-odd number oscillations in the probability distribution function.

\begin{figure}
	\includegraphics[width=1.05\linewidth,angle=0]{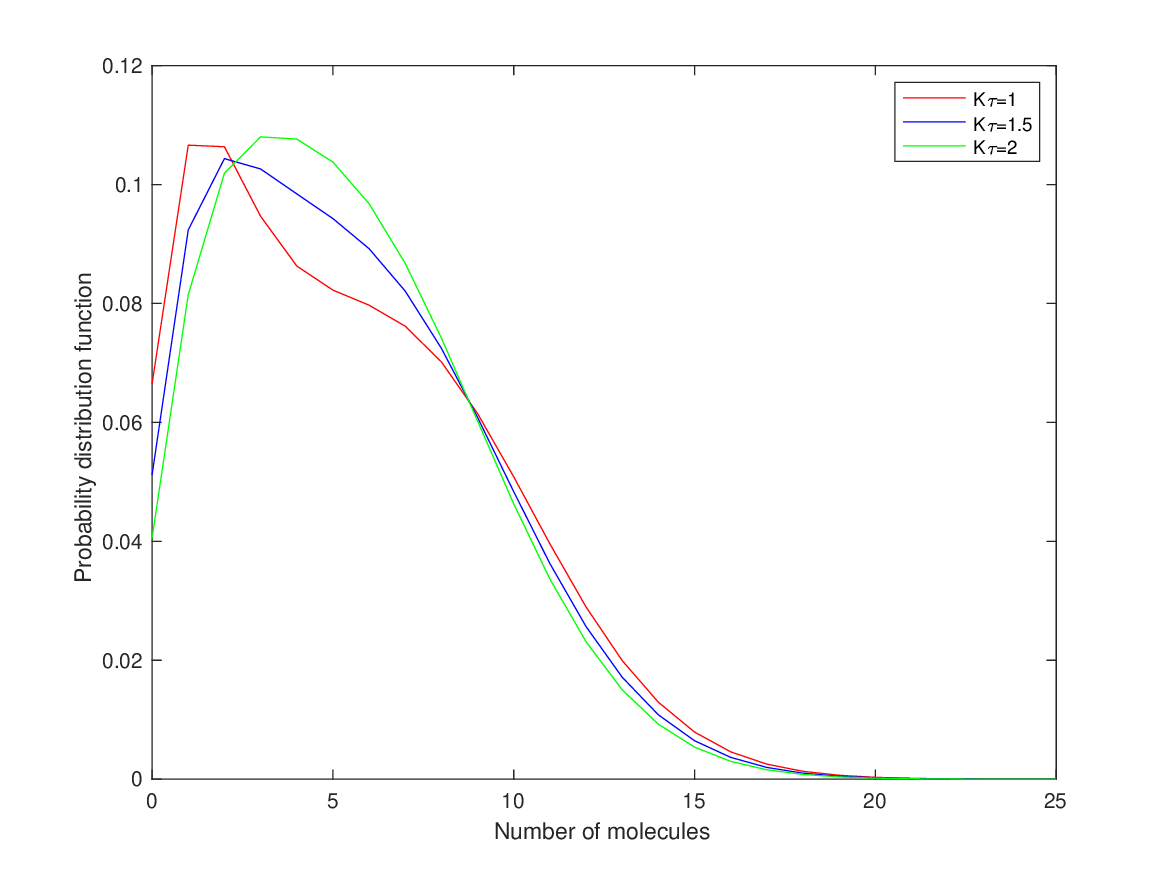}
	\caption{(Color online) Probability distributions for the following set of parameters: $k_0=k_1=0.5$, $\alpha_0=1$, $\alpha_1=10$.}
	\label{fig5}
\end{figure}

\begin{figure}
	\includegraphics[width=1.05\linewidth,angle=0]{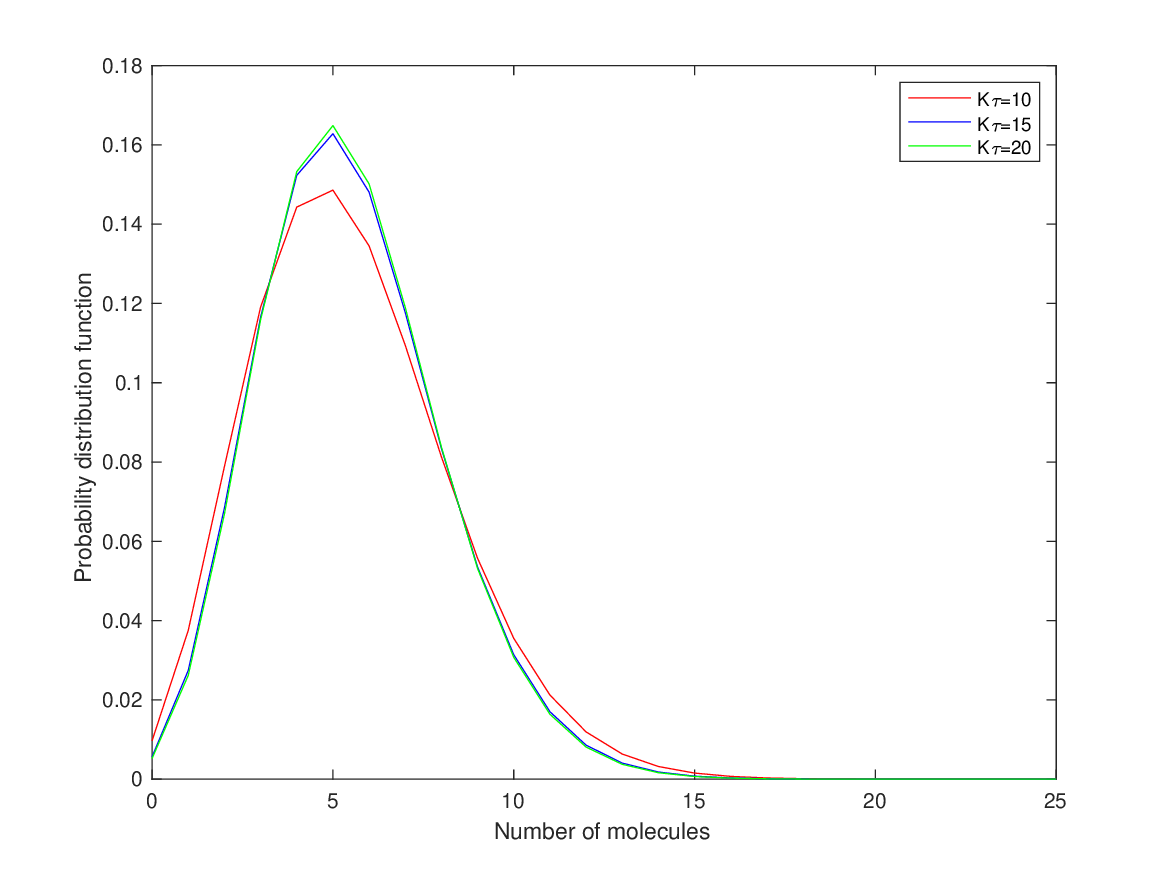}
	\caption{(Color online) Probability distributions for the following set of parameters: $k_0=k_1=0.5$, $\alpha_0=1$, $\alpha_1=10$.}
	\label{fig6}
\end{figure}

Let us look at the various regimes for the process of gene expression. 
The cases of relatively low production rates $\alpha_{0,1}$ and low rate $\alpha_0$ with high rates $\alpha_1$ are plotted in Figs.(\ref{fig3}) and (\ref{fig4}). 
Distributions for low rate $\alpha_0$ and high rate $\alpha_1$ for relatively low and relatively high $K$'s are presented in Figs.(\ref{fig5}) and (\ref{fig6}).
As the gene copy production rate $\alpha_1$ increases the distribution tends to become bimodal. Meanwhile we notice that increase in $K$, when the switching becomes faster,
the bimodality gets suppressed, and then disappears. For large $K$'s the distribution tends to a single-peaked function. We have calculated (\ref{pdf})  for
various set of parameters. The oscillations in the probability of even/odd number of molecules are absent. In the adiabatic case of large values of $K$
the oscillations are not just absent but the bimodal function becomes monomodal.

In order to understand the absence of the phenomenon of even-odd oscillations let us consider the adiabatic case analytically. First of all there is an error in the derivation of Eq.(\ref{adiame}) in \cite{kepler} (see Appendix).
At the same time there is a way to obtain the distribution (\ref{solution}). Let us look at the regime of large $K \gg 1$ when fast switching occurs. One can notice that the quasiprobability function (\ref{qpdf}) tends to become the Gaussian
\begin{eqnarray}
f(\alpha)=\frac{1}{\sqrt{2\pi\sigma^2}}\exp\left[-\frac{(\alpha-\bar{\alpha})^2}{2\sigma^2}\right]
\end{eqnarray}
centered at $\bar{\alpha}=(\alpha_0 k_1 + \alpha_1 k_0)\tau$ with the variance $\sigma^2=\frac{k_0k_1}{K}(\epsilon_1-\epsilon_0)^2$. In the limit $K\rightarrow\infty$, one has $f(\alpha)\rightarrow\delta(\alpha-\bar\alpha)$ and the probability distribution (\ref{pdf}) becomes Poissonian $P_n=\frac{\bar{\alpha}^n}{n!}e^{-\bar{\alpha}}$ with the average number of particles $\bar{\alpha}$. For large but finite $K$ the probability distribution is given by the expression (\ref{solution}) only if the 
integration limits in (\ref{pdf}) are taken $\pm \infty$
\begin{eqnarray}
P_n=\frac{C_1}{\sqrt{2\pi\sigma^2}}\int_{-\infty}^{+\infty} d\alpha \cdot \frac{\alpha^n}{n!}e^{-\alpha}\cdot \exp\left[-\frac{(\alpha-\bar{\alpha})^2}{2\sigma^2}\right] \\
= \frac{(i\sigma\sqrt{\frac{\tau}{2}})^n}{n!}H_n\left(i\sigma\sqrt{\frac{\tau}{2}}\left(1-\frac{\bar{\epsilon}}{\sigma^2\tau}\right)\right)\cdot\exp\left[-\bar{\epsilon}+\frac{1}{2}\sigma^2\tau\right] \nonumber
\end{eqnarray}

The even-odd number oscillations \cite{oscillations,pseudogenes,multistability} are well manifested in the limit $\bar{\alpha}\rightarrow\sigma^2\tau$. That corresponds to $k_1\epsilon_0+k_0\epsilon_1 \rightarrow \frac{k_0k_1}{K}(\epsilon_1-\epsilon_0)^2$. Let us recall that the adiabatic limit was taken requiring $k_1\epsilon_0+k_0\epsilon_1 \ll K$. These constrains can be satisfied when, for instance, $\epsilon_0$ and $k_0$ are both small quantities while $\frac{k_1\epsilon_1}{K} \rightarrow 1$.
However such approximation (taking integration limits $\pm \infty$ in (\ref{pdf})) would be incorrect. This is as rough approximation as the approach in \cite{kepler} (see Appendix) and actually describes another system. That is the model \cite{oscillations} 
with no switching but with a stochastic production rate.

{\it Master equation for (bio)chemical reactions with external noise. - }Let us now consider the following master equation
\begin{eqnarray}
\frac{d P^*_n}{dt}=\hat L P^*_n+\epsilon(P^*_{n-1}-P^*_n)
\label{me}
\end{eqnarray}
where $\hat L P^*_n$ being a regular right hand side of master equation (for probability distribution function $P^*_n$) that describes an arbitrary (bio)chemical reaction and $\epsilon$ the production rate. Under external noise we will assume fluctuations of $\epsilon=\bar\epsilon+\sqrt{2}\sigma\xi(t)$ with $\bar\epsilon$ being the mean value of $\epsilon$ and $<\xi(t)>=0$ , $<\xi(t)\xi(t')>=\delta(t-t')$; $\sigma$ is the intensity of the white noise $\xi(t)$. The equation for averaged out (over the external noise $\xi(t)$) probability distribution $P_n=<P^*_n>_{\xi}$ can be derived as follows. For $\textbf{P*}_n=(P^*_0, P^*_1, \hdots , P^*_{n-1}, P^*_n, P^*_{n+1}, \hdots)$ we have
\begin{equation}
\frac{d <\textbf{P*}_n>_{\xi}}{dt}> = A<\textbf{P*}_n>_{\xi} + B <\textbf{P*}_n \xi(t)>_{\xi}
\end{equation} 
where $A$ is the matrix derived from operator $\hat L$ plus the following matrix coming from $\bar\epsilon(\tilde P_{n-1}-\tilde P_n)$ term
\[\bar\epsilon \cdot
\begin{bmatrix}
-1	& 0	& 0	& 0 & \hdots \\
1 & -1 & 0 & 0  & \hdots  \\
0 & 1 & -1 & 0  & \hdots \\
0 & 0 & 1 & -1 & \hdots  \\
\hdotsfor{5}
\end{bmatrix}
\]
and matrix $B$ equals to
\[B=\sigma \sqrt{2} \cdot
\begin{bmatrix}
-1	& 0	& 0	& 0 & \hdots \\
1 & -1 & 0 & 0  & \hdots  \\
0 & 1 & -1 & 0  & \hdots \\
0 & 0 & 1 & -1 & \hdots  \\
\hdotsfor{5}
\end{bmatrix}
\]
Using a well-known formulas for the averaging procedure for linear multiplicative stochastic differential equations \cite{vankampen} we arrive at
\begin{equation}
\frac{d\textbf{P}_n}{dt}=A \textbf{P}_n+B_2 \textbf{P}_n
\label{vector}
\end{equation}
with
\[B_2=\sigma^2 \cdot
\begin{bmatrix}
1	& 0	& 0	& 0 & \hdots \\
-1 & 1 & 0 & 0  & \hdots  \\
1 & -2 & 1 & 0  & \hdots \\
0 & 1 & -2 & 1 & \hdots  \\
\hdotsfor{5}
\end{bmatrix}
\]

The scalar master equation that corresponds to the matrix master equation (\ref{vector}) would be
\begin{eqnarray}
\frac{dP_n}{dt}=\hat L_2 P_n+\sigma^2(P_{n-2}-2P_{n-1}+P_n)
\end{eqnarray}
where $\hat L_2 = \hat L + \epsilon(P^*_{n-1}-P^*_n)$.

{\it Relation to experimental and theoretical studies. - }There have been several experimental studies of stochastic gene expression processes among which there were oscillations measured \cite{skinner,skinner2}. In the paper \cite{skinner} in Fig. (3) one can see probability distributions go down and up. Although irregularly at first sight there could be a hidden mechanism behind such behaviour. There are a few downs and ups in Figs. (1) and (3) in \cite{skinner2}. The paper \cite{yanagida} has Figs.(4h)-(4j) as well as Fig.(5d) that show quite regular downs and ups. The authors of \cite{yanagida} call it "Poisson with zero spike" meaning that there is non-zero probability for having no molecules plus a Poisson distribution with non-zero mean value. However such distribution does not describe those downs and ups. Although it remains to make a more detailed quantitative comparison with the experimental data a qualitative picture emerges.

Let us point out that the analysis we presented can be applied to switching processes that occur between two steady states in the linear approximation. That means that one can start with a model for transcriptional regulation with a feedback that would involve nonlinearities and then linearize the system around steady states thus reducing to the linear inter-state switching. Although we had the same relaxation rate $\tau^{-1}$ for both states we expect that the results will not differ qualitatively. In order to get an idea of transcriptional regulation with a feedback one can look at the models of genes being self-regulated \cite{assaf,biancalani,assaf2} via, e.g., DNA (un)looping \cite{vilar saiz, loops, earnest}. We have also recently considered a model of self-regulated genes and pseudogenes \cite{multistability} where the phenomenon of noise-induced multistability was desribed. Our analysis presented above suggests that switching alone cannot cause even-odd oscillations in probability distributions for numbers of particles. Thus addition of external noise in the production rate is essential. 

In order to see why the oscillations occur in the case of stochastic production rate with no switching let us write down the master equation for the case of equal production mean rate and noise intensity $\bar \epsilon = \sigma^2$
\begin{eqnarray}
\frac{dP_n}{dt}=\bar\epsilon (P_{n-1}-P_n) +\frac{1}{\tau}[(n+1)P_{n+1}-nP_n] \nonumber \\
+ \sigma^2(P_{n-2}-2P_{n-1}+P_n) \nonumber \\
= \bar\epsilon(P_{n-2}-P_{n-1}) + \frac{1}{\tau}[(n+1)P_{n+1}-nP_n] \nonumber
\end{eqnarray}
The equation implies that particles are being created in pairs. In order to get a non-zero $P_1$ one has to have non-zero $P_{-1}$ which is never the case as $P_{-1}=0$. As a result the probabilities of all odd numbers $P_{2n+1}=0$. 
This is the case shown in Fig.(\ref{fig1}). This meachnism is absent in the case of pure switching. Even in the case of fast transitions between states the system adjusts by switching to a different production rate.

In a recent publication \cite{larsson} the authors presented measurements for stochastic gene expression. In the section called "Extended Data" Figs. (4), (7), and (8) display the above mentioned downs and ups in the distributions of numbers of mRNA molecules. They studied a transcriptional burst kinetics that is a stochastic switching process (see \cite{bressloff} and references therein). Once again, those regular or irregular ups and downs cannot be described simply by the switching. In the theoretical model presented in our paper one has to introduce some noise, e.g., in the production rate. More research is necessary to see whether a combination of stochastic processes, such as switching, external noise or other fluctuating parameters would be able to describe the outcomes of similar experiments.

Besides its biological significance \cite{oscillations,multistability} it is worth to mention that the oscillatory behaviour has a direct analogue in quantum physics. There exist oscillations in photon number distributions \cite{schleich} which are caused by superposition of coherent states of light with quantum noise \cite{perina} (see also \cite{dodonov} for a related review with a number of citations). In the case of (bio)chemical reactions the oscillations are induced by the convolution of the Poisson (due to the (bio)chemical reaction noise) and the Gaussian (due to the extrinsic intracellular noise) stochastic processes. That is especially clearly seen when one looks at both Poisson representation for the probability distribution function governed by the master equation and Glauber-Sudarshan or either positive or complex P-representation for the density matrix in quantum physics (see Gardiner's book \cite{gardiner} that contains a detailed description of both approaches).

Summing up, the adiabatic master equation obtained by Kepler and Elston \cite{kepler} (although using a rough approximation) contains an additional term that causes diffusion in particle number centered on $n-1$ rather than $n$. Meanwhile, the addition of extrinsic noise to the production rate in the absence of switching leads to the same qualitative result. This suggests that the master equation in the form identical to that of (\ref{adiame})
\begin{eqnarray}
\frac{dP_n}{dt}=\hat L P_n+\sigma^2(P_{n-2}-2P_{n-1}+P_n)
\end{eqnarray}
with $\hat L P_n$ being the conventional (standard) part of the (bio)chemical master equation, would provide a more complete description of (bio)chemical reactions, especially in the field of stochastic gene expression. The second main result is the absence
of even-odd oscillations in probability distributions for numbers of molecules in the case of inter-state switching. Our next goal will be to combine these two different models: the inter-state switching and molecule production with external noise. 

This work was supported by the Ministry of Science and Technology (MOST) in Taiwan under the grants MOST 106-2811-M-001-086, 107-2811-M-259-514 and 108-2112-M-259-008. 

{\it Appendix. - }The adiabatic approximation in the paper \cite{kepler} is presented in Eqs. (19)-(21). In our notations the equations for $P_n=P_n^0+P_n^1$ and $\zeta=k_0P_n^0-k_1P_n^1$ are as following
\begin{eqnarray}
\dot P_n=(\epsilon_0 k_1 + \epsilon_1 k_0)(P_{n-1}-P_n) \nonumber \\
+\frac{1}{\tau}[(n+1)P_{n+1}-nP_n] + (\epsilon_0-\epsilon_1)(\zeta_{n-1}-\zeta_n) 
\label{eq1}
\end{eqnarray}
\begin{eqnarray}
\dot \zeta_n = -K\zeta_n + (\epsilon_0 k_1 + \epsilon_1 k_0)(\zeta_{n-1}-\zeta_n) \nonumber \\
+\frac{1}{\tau}[(n+1)\zeta_{n+1}-n\zeta_n]+k_0k_1(\epsilon_0-\epsilon_1)(P_{n-1}-P_n)
\label{eq2}
\end{eqnarray}
While Eq.(\ref{eq1}) coincides with Eq.(19) in \cite{kepler}, Eq.(\ref{eq2}) differs from Eq.(20) derived in \cite{kepler}. There are extra factors $(k_0-k_1)$ and $(k_0-k_1)^2$ in the second and third terms in the right hand side of the Eq.(20)
as well as the sign is opposite in the  fourth term. In order to obtain Eq.(21) one has to drop the second and third terms in the right hand side of both Eq. (20) in \cite{kepler} and Eq.(\ref{eq2}), keep the correct fourth term in Eq.(\ref{eq2}) as well as assume $\dot \zeta = 0$ (adiabatic elimination of the fast variable). The final result would be Eq.(\ref{adiame}).

\end{document}